# Strong variability of Martian water ice clouds during dust storms revealed from ExoMars Trace Gas Orbiter/NOMAD

**Giuliano Liuzzi[1,2], Geronimo L. Villanueva[1], Matteo M. J. Crismani[3], Michael D. Smith[1], Michael J. Mumma[1], Frank Daerden[4], Shohei Aoki[4], Ann Carine Vandaele[4], R. Todd Clancy[5], Justin Erwin[4], Ian Thomas[4], Bojan Ristic[4], José-Juan Lopez-Moreno[6], Giancarlo Bellucci[7], Manish R. Patel[8,9]**

[1] NASA Goddard Space Flight Center, 8800 Greenbelt Rd., Greenbelt, 20771 MD, USA

[2] Dep. of Physics, American University, 4400 Massachusetts Av., Washington, 20016 DC, USA

[3] NPP/USRA, NASA Goddard Space Flight Center, Planetary Systems Laboratory, Code 693, Greenbelt, MD, USA

[4] Royal Belgian Institute for Space Aeronomy, BIRA-IASB, 3 Avenue Circulaire, 1180 Brussels, Belgium

[5] Space Science Institute, 4750 Walnut Street, Suite 205, Boulder, CO 80301, USA

[6] Instituto de Astrofisica de Andalucia, IAA-CSIC, Glorieta de la Astronomia, 18008 Granada, Spain

[7] Istituto di Astrofisica e Planetologia Spaziali, IAPS-INAF, Via del Fosso del Cavaliere 100, 00133 Rome, Italy

[8] School of Physical Sciences, The Open University, Milton Keynes, MK7 6AA, UK

[9] SSTD, STFC Rutherford Appleton Laboratory, Chilton, Oxfordshire OX11 0QX, UK

Corresponding author: Giuliano Liuzzi (giuliano.liuzzi@nasa.gov)

**Key Points:**

- Water vapor condensation rapidly responds to dust storms, increasing the altitude of mesospheric cloud formation.

- Clouds structure is observed to change between dawn and dusk, indicating nighttime nucleation of water ice on dust particles.

- Vertical gradients in water ice particle size relate to water vapor and nuclei availability, which vary between the two dust storms of MY34.

## Abstract

Observations of water ice clouds and aerosols on Mars can provide important insights into the complexity of the water cycle. Recent observations have indicated an important link between dust activity and the water cycle, as intense dust activity can significantly raise the hygropause, and subsequently increase the escape of water after dissociation in the upper atmosphere. Here present observations from NOMAD/TGO that investigate the variation of water ice clouds in the perihelion season of Mars Year 34 (April 2018-19), their diurnal and seasonal behavior, and the vertical structure and microphysical properties of water ice and dust. These observations reveal the recurrent presence of a layer of mesospheric water ice clouds subsequent to the 2018 Global Dust Storm. We show that this layer rose from 45 to 80 km in altitude on a timescale of days from heating in the lower atmosphere due to the storm. In addition, we demonstrate that there is a strong dawn dusk asymmetry in water ice abundance, related to nighttime nucleation and subsequent daytime sublimation. Water ice particle sizes are retrieved consistently and exhibit sharp vertical gradients (from 0.1 to 4.0 μm), as well as mesospheric differences between the Global Dust Storm (<0.5 μm) and the 2019 regional dust storm (1.0 μm), which suggests differing water ice nucleation efficiencies. These results form the basis to advance our understanding of mesospheric water ice clouds on Mars, and further constrain the interactions between water ice and dust in the middle atmosphere.

## 1 Introduction

Aerosols are an essential component of the atmospheric energy budget of Mars, and their constant presence significantly effects the climate, radiative equilibrium and meteorology of the planet (e.g. *Heavens et al., [2011]; Määttänen et al., [2013]; Madeleine et al., [2011]; Whiteway et al., [2009], Haberle et al., [2017]*). Studying this aspect of Mars requires knowledge of water ice and dust microphysical properties, spatial distribution, and temporal variability, which has benefited from observational data from several instruments on various missions. A comprehensive picture of Martian water ice and dust, and of their temporal and spatial variability was presented in *Smith [2004, 2008]* using the Thermal Emission Spectrometer on Mars Global Surveyor (MGS-TES, *Christensen et al. [2001]*). These results showed that the perihelion season ($L_S = 180\text{-}360$) is relatively warm due to the enhanced solar flux, has pronounced dust activity and limited water ice cloud column opacities. On the other hand, the aphelion season ($L_S = 0\text{-}180$) is generally cooler, has low dust activity and inter-annual variability, with the formation of thick water ice clouds within the hygropause in the tropical region.

Recent work (e.g. *Heavens et al., [2011]; Kleinböhl et al., [2009]; Sefton-Nash et al., [2013] Guzewich et al. [2014]; Smith et al. [2013]; Wolff et al. [2009]*; *Clancy et al. [2019]; Guzewich & Smith [2019]*) constrains the vertical distribution of water ice and dust using data acquired in limb geometry by the Compact Reconnaissance Imaging Spectrometer for Mars (CRISM, *Murchie et al., [2007]*), the Mars Climate Sounder (MCS, *McCleese et al., [2007]*) on the Mars Reconnaissance Orbiter (MRO), and UV solar occultations by SPICAM (Spectroscopy for Investigation of Characteristics of the Atmosphere of Mars) *[Bertaux et al., 2006, Montmessin et al., 2017]* on Mars Express. These data provide substantial evidence for complexity in the vertical distribution of water ice and dust particles, showing that water ice clouds can be found at altitudes up to 80 km during perihelion season, and that dust is infrequently found above 60 km.

Moreover, it has been shown that water ice particles have a vertical gradient in their size distribution *[Fedorova et al., 2014; Clancy et al., 2019]*, with larger particles (1 to 5 μm) found in the lower atmosphere, and much smaller particles (0.1 to 0.6 μm) in the mesosphere. CRISM data also suggests that particle size is seasonally dependent *[Guzewich & Smith, 2019]*, which may offer hints about the dynamics and timescale of the processes that drive cloud formation and transport through the middle atmosphere.

Despite detailed studies that span a range of temporal and spatial coverage, the current description of Martian water ice and dust is incomplete. Formation mechanisms of ice clouds and interactions with dust and radiation were initially investigated in the theoretical works of *Michelangeli et al. [1993]* and *Rodin et al. [1999]*. Recent works *[Neary et al, 2019; Heavens et al, 2018; Navarro et al., 2014]* have gained detailed insights into modeling the feedback between the dust and water cycles, and the mechanisms by which a dust storm can expand on a global scale and affect atmospheric circulation. Previous observations have shown that elevated, optically thick layers of dust in the atmosphere increase the temperature of the surrounding air by tens of degrees up to 60 km (see e.g. *Gurwell et al. [2005]*). This behavior was characterized over many Martian Years (MYs), during dust storms, and in comparison with TES mm-wave studies *[Clancy et al., 2000]*, including an otherwise unobserved perihelion dust storm in 1994 (MY21). This feature of the Martian atmosphere challenged the Viking characterization of a 15K unusually warm aphelion atmosphere (that incorrectly precluded the existence of the Aphelion Cloud Belt *[Clancy et al., 1996]*). Therefore only recent modeling and observations have demonstrated the effect this has on water vapor circulation, causing it to lift in the atmosphere up to 70 km (*Heavens et al. [2018]; Vandaele et al. [2019]*) on a Sol timescale.

Current circulation models and databases (e.g. Mars Climate Database (MCD), *Millour et al., [2015]*) can simulate a range of cloud formation processes and predict seasonal cloud morphologies and formation altitudes with general agreement with observations. However, there are some open questions still unsolved: in particular, the gradient in the particle size of water ice crystals between the troposphere and mesosphere (e.g. *[Guzewich et al., 2014]*), is still challenging to explain, since small particles (~0.1 μm) are unstable against coagulation *[Fedorova et al., 2014]*. While estimating the importance of coagulation and sedimentation requires a precise knowledge of the number density of coagulating particles *[Michelangeli et al., 1993]*, their frequent observation implies a stable source of particles. Some have posited that dust could be lifted from the surface *[Spiga et al. 2013]*, yet the efficiency of this mechanism in triggering water ice nucleation in the mesosphere at perihelion needs to be assessed. An additional hypothesis has been suggested due to the recent discovery of ablated meteoric dust *[Crismani et al., 2017]*, which forms meteoric smoke particles that may act as efficient nucleation sites *[Plane et al., 2018, Hartwick et al., 2019]*.

Data acquired during the MY34 perihelion season can offer important constraints about these novel hypotheses, at least during the perihelion season. Although this work does not directly assess the role of coagulation and sedimentation, it provides a large casuistry of water ice clouds observation and their properties from which further modeling work can assess the verity of their assumed microphysical models. In general, comprehensive data analyses that investigate the interactions of water ice and dust from the troposphere to the mesosphere on a global scale are still rare.

Here we show a novel characterization of water ice and dust vertical distribution and properties, obtained using the data acquired by the Nadir and Occultation for MArs Discovery (NOMAD) instrument of the ExoMars Trace Gas Orbiter (TGO) spacecraft *[Neefs et al., 2015; Vandaele et al., 2018; Vandaele et al., 2015]*. These results include the entire perihelion season, between April 2018 ($L_S$ = 162.5, MY 34) and April 2019 ($L_S$ = 15.0, MY 35) and altitudes up to 110 km. Such vertical coverage goes beyond the value of 80 km previously reached by SPICAM *[Fedorova et al., 2018]*. Besides trace gases, NOMAD data are suitable to retrieve aerosol properties. NOMAD was designed to observe primarily in Nadir and Solar Occultation geometry. The latter probes the atmosphere vertically from the surface to high altitudes with high signal-to-noise ratios (SNR). Similarly to SPICAM solar occultations, this permits retrievals of vertical profiles of atmospheric aerosols closer to the surface of the planet (5-10 km) than generally obtained by limb radiance observations (e.g. CRISM and MCS), which typically support vertical retrievals down to ~15-30 km altitudes subject to atmospheric aerosol loading. Similar to CRISM and MCS, the spectral interval observed by NOMAD (2.2 to 4.3 μm), contains clear signatures of water ice extinction, while dust absorption/scattering is mostly a continuum. Still, this allows NOMAD to distinguish the two aerosols in the large majority of the cases. Moreover, since measurements are acquired at both Mars' terminators (dawn and dusk) on a daily basis (up to 12 occultations per Sol), NOMAD retrievals provide information about the seasonal behavior of aerosols in the atmosphere with altitude, latitude, and local time (dawn and dusk). Although TGO solar occultations do not obtain contiguous latitudinal coverage daily, they nevertheless provide information about the global properties of aerosols at the terminator. Particular attention has been dedicated to evaluate the impact of the 2018 Global Dust Storm (GDS) on water ice clouds formation and dissipation.

The manuscript is structured as follows: in Section 2 we provide details about the NOMAD instrument and observations, and how they have been used to retrieve aerosols concentration and properties. In Sections 3 and 4 we describe the results, and discuss the temporal and spatial variability of aerosols. Conclusions are drawn at in Section 5.

## 2 The NOMAD instrument and data structure

### 2.1 The NOMAD instrument

The Nadir and Occultation for MArs Discovery (NOMAD, *Vandaele et al. [2018]; Vandaele et al. [2015]*) instrument is onboard the ExoMars Trace Gas Orbiter 2016, an ESA/NASA mission to Mars. The spacecraft comprises also another multi-channel spectrometer, the Atmospheric Chemistry Suite (ACS, *[Korablev et al., 2017]*). The main objective of NOMAD and ACS is to investigate Mars' atmosphere at unprecedented spectral resolution in the UV, Visible, and IR. NOMAD operates between 0.2-0.65 and 2.2-4.3 μm, and is a compact, high-resolution instrument composed of 3 channels: a solar occultation channel (SO) that operates in the IR, a second infrared channel, LNO, mostly used for nadir measurements (but capable also of limb observations), and an ultraviolet/visible channel (UVIS) that can work in all observation geometries.

The design of SO and LNO channels is based on the SOIR spectrometer *[Nevejans et al., 2006]* that was developed for the ESA Venus Express mission. The infrared channels of NOMAD have been described in detail by *Neefs et al. [2015]* and *Thomas et al. [2016]*, and here we recall only their main features. Besides, *Liuzzi et al. [2019]* discusses the full calibration that used the in-flight data acquired prior to the science phase. ExoMars TGO is in a near-polar orbit, so NOMAD can ideally perform two occultations per orbit (24 occultations per Sol). However, this number reduces to 12-14 per Sol, due to constraints related to the operability of all the instruments onboard TGO.

This work focuses on the data acquired by the SO channel, whose routine science operations started in April 2018 and are ongoing. The SO channel works at wavelengths between 2.2 and 4.3 µm (2325 – 4500 cm$^{-1}$), and is based on an echelle grating in a Littrow configuration combined with an Acousto-Optic Tunable Filter (AOTF) for spectral window selection. The AOTF is a narrow bandpass filter, whose properties are tuned by a suitable input Radio Frequency that selects the central frequency where the AOTF transfer function peaks, limiting the signal that enters the spectrometer to a specific diffraction order. An order width varies between 20 and 35 cm$^{-1}$, increasing linearly with the diffraction order number. As the SO channel is pointed towards the Sun, it observes the solar radiation successively attenuated by the Martian atmosphere at increasing/decreasing altitudes (dawn/dusk), enabling investigation of the vertical structure of the atmosphere.

By switching diffraction orders with the AOTF, each measurement consists of repeated cycles of (typically) five or six key diffraction orders, whose spectral ranges include absorption by gases/organics of interest. Therefore, at each altitude the instrument produces observations in five to six narrow spectral intervals, providing effective broadband information covering the whole spectral range of NOMAD. Constraints in data transmission, and the need to increase the SNR, limit the number of output rows (effective or binned) to 24, as previously done by SOIR *[Mahieux et al., 2008]*. The 24 rows of the detector typically have an Instantaneous Field of View (IFOV) of 7.5 km in tangent heights. Hence, a single detector row will sample a vertical distance of 500 m. For routine science operations, data are usually transmitted to Earth in 4 bins, and each of them will contain the signal of 4 co-added rows of the detector. By correcting the IFOV by the orbital velocity and the angle between the line of sight and the surface of Mars, the effective FOV (resolution) of each bin will vary between 0.6 and 1.5 km.

## 2.2 Data and pre-processing

The radiation flux observed by NOMAD is the result of a complex convolution between the flux reaching the instrument, and the spectral functions describing both the AOTF transmittance and the grating properties (i.e. the Blaze function, *Liuzzi et al. [2019]*). Therefore the data need to undergo pre-processing before being used for retrieval purposes. When observing in Solar Occultation geometry, the initial step is the derivation of the transmittance for each observed order and each altitude from surface to the top of atmosphere (TOA). Spectra acquired by NOMAD at altitudes at which atmospheric extinction is negligible, above the TOA, provide unity transmission spectra at wavelength λ, $F(\lambda)$. Each atmospheric transmission spectrum, $I(\lambda)$, is divided by an average of a subset of these "reference" spectra, $F(\lambda)$. Once both $I(\lambda)$ and $F(\lambda)$ signal intensities are expressed in observed photons (derived from the data in ADUs knowing the detector quantum efficiency), the calculation of transmittance $T(\lambda)$ and related uncertainty $\sigma_T$ is straightforward:

$$T(\lambda) = \frac{I(\lambda)}{F(\lambda)} \tag{1}$$

$$\sigma_T(\lambda) = T(\lambda) \sqrt{\left(\frac{\sigma_I(\lambda)}{I(\lambda)}\right)^2 + \left(\frac{\sigma_F(\lambda)}{F(\lambda)}\right)^2} \tag{2}$$

Since in SO geometry the signal coming from the source is much larger than any thermal instrument background signal, it is reasonable to assume that the shot noise is dominant over other noise sources. Within this hypothesis, the noise behaves according to a Poisson distribution, and the uncertainties $\sigma_I$ and $\sigma_F$ are simply the square root of the signal:

$$\sigma_I(\lambda) = \sqrt{I(\lambda)}; \ \sigma_F(\lambda) = \frac{1}{\sqrt{N}} \sqrt{\sum_{j=1,\dots,N} F_j(\lambda)} \tag{3}$$

with $N$ the number of reference spectra.

The choice of the subset of reference spectra $\{F_j(\lambda)\}_{j=1,\dots,N}$ is critical to maximize the quality of the derived transmittances. In fact, the instrument is affected by temperature variations even along an occultation, that cause two different effects. The first results in micro-misalignments between the wavenumber scales of reference spectra and the observed spectrum *[Liuzzi et al., 2019]*. This produces dispersion features in the derived transmittance, in correspondence of solar lines, which can be comparable or larger than atmospheric absorption features, especially at high altitudes, and can potentially significantly degrade the SNR corresponding to the uncertainties in Eq. ( **3** ). The second effect is the shift of the spectral response function caused by the shift of the AOTF function center, which is related to thermally induced micro-deformations of the AOTF crystal itself. This typically results in artifacts in the transmittance continuum, producing fluctuations around its average value as large as 1%. The impact of such systematics on retrievals will be shown in Section 3.2.

To mitigate these issues, the altitude at which the set of reference spectra is located is always chosen as closest as possible to the TOA, and the number of reference spectra to average in Eq. ( **3** ) is limited to make them fall in an altitude range no larger than 20 km. This is done in the realistic assumption (verified on the data) that the instrument temperature varies slowly compared to the acquisition time of the six orders typically measured during an occultation (~1 second).

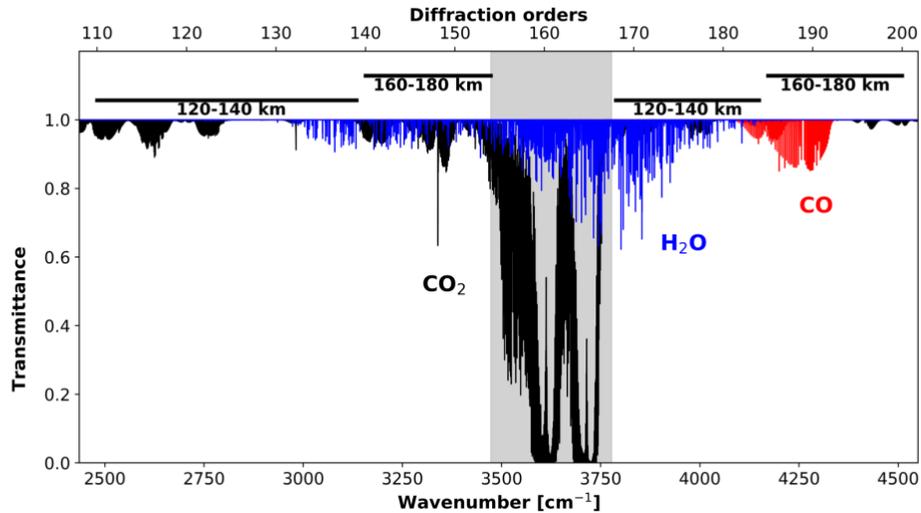

**Figure 1**. *Representation of an average Martian atmospheric transmittance (aerosol free) in Solar Occultation at h=20 km. Absorption by $CO_2$, $H_2O$ and $CO$ is highlighted. The top panel reports the position of NOMAD SO diffraction orders and the respective TOA height. TOA is related to order opacity. The grey zone corresponds to the spectral interval (and the orders) not used for aerosol retrieval, because of strong $CO_2$ absorption.*

The definition of the TOA is not the same for each order. For the sake of normalization, the TOA depends on the altitude at which atmospheric extinction is no longer detected, which depends strictly on the wavenumber at which each diffraction order is centered. This concept is illustrated in Figure 1, where we report a summary of the spectral absorptions by gases and aerosols in the Martian atmosphere in the spectral interval covered by NOMAD. Based on these, different TOAs have been adopted in differing spectral intervals (top panel of Figure 1). Because of the strong $CO_2$ absorption, the spectra acquired at diffraction orders 155 to 167 are not used for aerosol retrievals. If six diffraction orders are measured in an occultation, one of them is always in the interval 155-167, therefore all aerosol retrievals use five orders.

The spectra used in this study correspond to data Level 0.3, version 1.0 of the data provided by the NOMAD PI institute (nomad.aeronomie.be).

## 2.3. Retrieval methodology

In this work, NOMAD data are used to retrieve water ice and dust abundance, simultaneously with their particle radii. As previously noted, NOMAD measures five or six diffraction orders during a typical occultation. Besides those heavily affected by $CO_2$ saturation (see Figure 1), all of the spectra acquired in other diffraction orders show well-separated absorption lines, and a continuum transmittance that provides information about aerosol extinction. While the radiation is mostly extinct by the aerosols through scattering, the contribution of multiple scattering processes is negligible, because the observed flux is highly directional (the used detector rows are fully illuminated by the Sun disk). This assumption has been verified prior to retrievals, performing sample retrievals on selected NOMAD data.

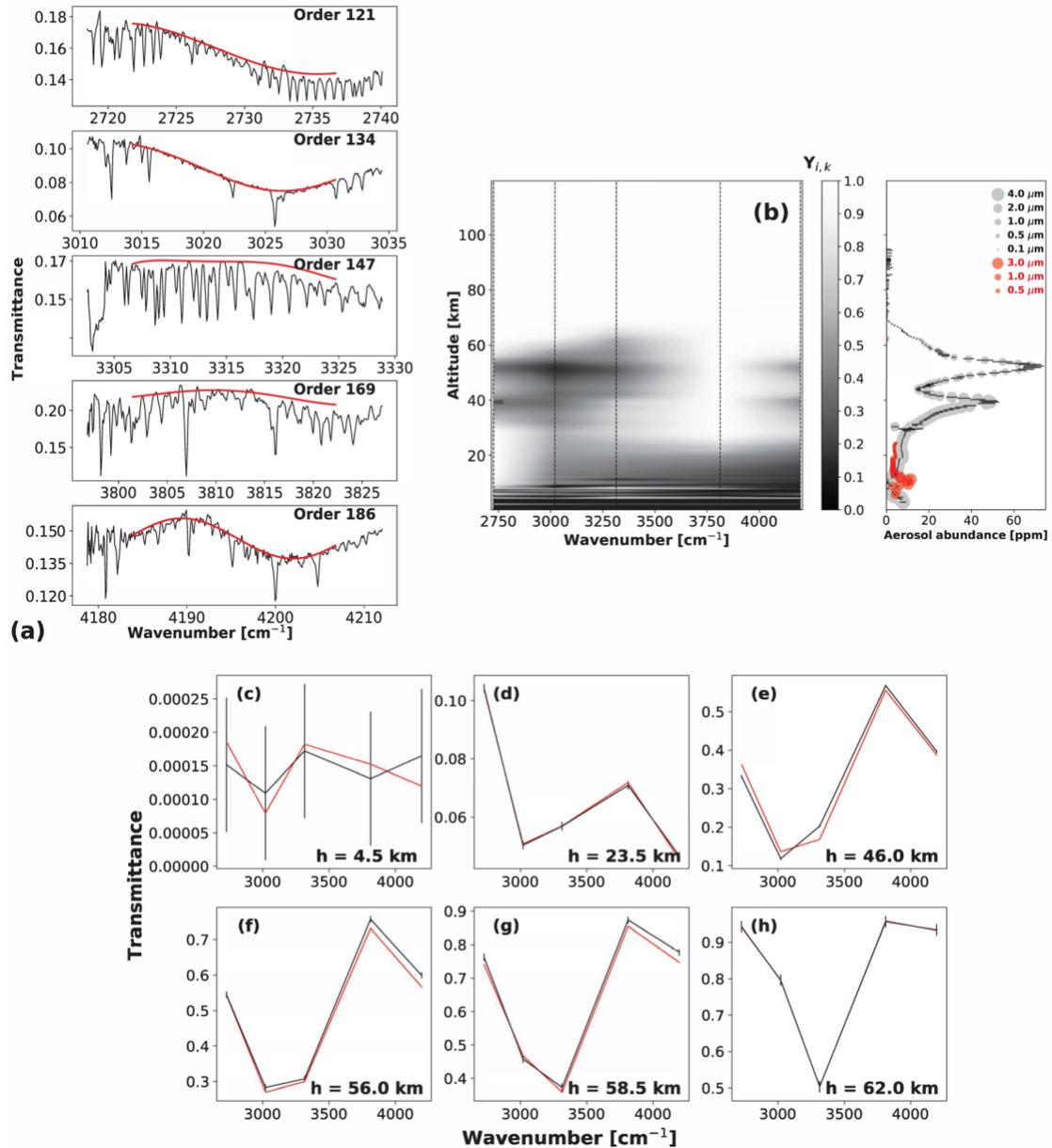

**Figure 2**. *Derivation of transmittance during an occultation. Actual NOMAD data are presented. a) data from 5 diffraction orders taken at each altitude (black spectra, at h=30 km). The red line is the polynomial fit to the continuum in the central 220 pixels; b) Relative depth, layer by layer, of the average polynomial continuum, interpolated in altitude and wavenumbers. The vertical lines represent the position of the 5 orders on the left. The retrieved abundances and particle sizes of water ice (black) and dust (red) are on the right. Transmittance drops below 1% at 10 km; c) to h): representation of the broadband transmittances (Y) at different altitudes. No significant information is found at low altitudes (c). Red lines are the best fits (in (h) superimposed to data). Data is from 2018 Oct 23, lat -20 to -25, lon 75 to 80 E (north of Hellas Basin).*

Aerosol concentrations and radii are retrieved for a set $\{h_i\}_{i=1,...,H}$ of $H$ discrete points in altitude, which vary occultation by occultation. For each $h_i$, 5 (one per order) are available. Any occultation with less than 5 orders that are suitable for aerosol retrieval (outside the interval 155-167) are not considered, such that there is sufficient information content available for the retrieval process. Depending on the orders themselves, and the wavelength at which they are centered, the overall information content in terms of aerosol properties can vary.

For each spectrum $\{T_{il}\}_{i=1,...,H \atop l=1,...,M}$, where $M$=320 is the number of spectral points, we derive the average "continuum" transmittance (free of gas absorption) by fitting a 5$^{th}$ degree polynomial to the spectral continuum, and taking its average value. While Eq. (3) holds in the hypothesis of pure shot noise, it is necessary to associate to each continuum average an uncertainty reflecting the systematics caused by fluctuations of the instrument temperature. Therefore, to maximize the signal-to-noise ratio, we fit only the continuum of the central $M_0 = 220$ points (excluding the first and last 50 points), since the signal on the edges is significantly suppressed by the grating's blaze function, and the corresponding uncertainty is calculated as the standard deviation of the $M_0$ continuum points considered.

Once the average transmittance is calculated for each order and altitude, the input data for the retrieval process is constituted by a matrix $\{Y_{ik}\}_{i=1,...,H \atop k=1,...,5}$, with $k$ index indicating diffraction orders. An example of NOMAD data is presented in Figure 2, where the process of deriving the matrix of data $Y$ is also shown. The data are presented in the form of a continuous contour plot, which is useful to demonstrate the variation of the continuum properties with altitude. At each altitude sampled by NOMAD there will be a 5-point spectrum and some examples are shown in the bottom panels. It can be seen that the data demonstrate signatures of water ice concentration and particle size, and their variation with altitude. Dust scattering is instead visible throughout the entire spectral interval of NOMAD.

Figure 3 emphasizes the information content of NOMAD data for aerosols, illustrating a series of sample calculations of transmittances ($h = 40$ km), obtained by individually varying the four parameters we retrieve. NOMAD data in the form of matrix $Y$ show distinct trends when water ice and dust concentration increase; for water ice size, the data can discriminate between sub-micron and larger particles, with enhanced sensitivity to variations in the sub-micron domain. For dust, we expect the information about particle size to be partially entangled with concentration, as expected from previous studies in comparable spectral intervals (e.g. *Smith et al., [2013]*).

We have precisely quantified values, uncertainties, and the independence between the retrieved parameters through a customized retrieval procedure optimized for this data. Each broadband spectrum $\{Y_{ik}\}_{i=i0 \atop k=1,...,5}$ is analyzed through a forward radiative transfer model and a retrieval procedure.

Radiative transfer and retrievals are performed using the Planetary Spectrum Generator (PSG, *Villanueva et al., [2018]*). PSG is an online tool capable of synthesizing planetary spectra (atmospheres and surfaces) in a broad range of wavelengths (0.1 μm to 100 mm) from any observatory (e.g., ALMA, JWST), orbiter (e.g., ExoMars TGO, Cassini), or lander (e.g., MSL,

InSight). This is accomplished by combining several state-of-art radiative transfer models, spectroscopic databases, and planetary circulation and climatological models.

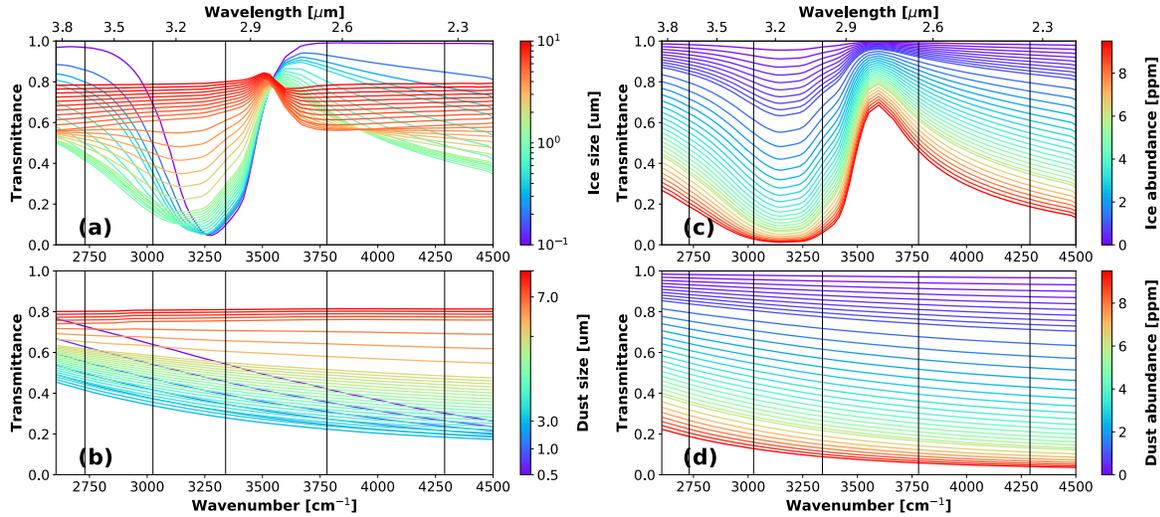

**Figure 3**. *Simulation of the effect on the observed transmittance due to variations in: (a) water ice particle size; (b): dust particle size; (c): water ice abundance; (d): dust abundance. Spectra correspond to the slant path transmittance at a tangent altitude of 40 km, with a nominal dust/ice concentration of 5 ppm (left) and a size of 1 μm (right), constant with altitude.*

In this work, we calculate the opacity due only to the aerosol component since the data are representative of the spectra continuum and the gas absorption is unresolved; hence we do not use the full radiative transfer capabilities of PSG. For this case study, calculations are made using a pseudo-spherical approximation (e.g. *Smith et al. [2013]*) implemented by PSG; this is essential to correctly account for the integrated column along the line of sight. In general, the model is also able to perform multiple scattering from atmospheric aerosols, using the discrete ordinates method (*Stamnes et al. [2017]*).

PSG includes optical constants for more than 100 species of aerosols. For dust, we rely on the refractive indices derived with CRISM data *[Wolff et al., 2009]*, to compute the extinction for concentration and particle sizes, which here are intended as particle radius. Water ice opacity is computed based on the spectral properties provided in *Warren & Brandt [2008]*. Both for water ice and dust, we assume a log-normal size distribution with a variance equal to 0.18 μm.

The spectral resolution of a single broadband spectrum is equal to the Free Spectral Range (FSR) of NOMAD by construction, namely the interval between the center wavenumber of two consecutive orders, which is a constant of the instrument and is equal to 22.55 cm$^{-1}$ *[Liuzzi et al., 2019; Neefs et al., 2015]*. Therefore, for the sake of retrievals, spectra are computed with a resolution equal to the FSR, between 2000 and 5000 cm$^{-1}$, and internally interpolated to the 5 spectral points of each broadband spectrum.

As mentioned, retrievals quantify the properties of aerosols, while the thermal structure of the atmosphere is assumed. For each spectrum, PSG calculates an a-priori atmospheric status by collocating the climatology provided by the GEM model *[Neary et al., 2019]*, which is tuned on the MY34 GDS scenario, consistently with the observations. This is crucial because aerosol concentration is retrieved in terms of mass mixing ratio, which is highly sensitive to the actual density of the observed atmosphere, for which an accurate reference is therefore necessary. This aspect will be further discussed upon presenting retrieval results (Section 3.2).

Retrievals are based on Optimal Estimation (OE, *Rodgers [2000]*), where the optimal solution is sought through a Gauss-Newton iterative approach, and each spectrum is analyzed individually. Given the observed spectrum $\boldsymbol{R}$, and $F$ the function representing the radiative transfer model, we first define the Jacobian matrix $\boldsymbol{K} = \frac{\partial F(\boldsymbol{v})}{\partial \boldsymbol{v}}\Big|_{\boldsymbol{v}=\boldsymbol{v}_0}$, where $\boldsymbol{v}$ is the parameters vector, and $\boldsymbol{v}_0$ is the parameters' initial guess. If we also define $\boldsymbol{x} = \boldsymbol{v} - \boldsymbol{v}_0$ and $\boldsymbol{y} = \boldsymbol{R} - F(\boldsymbol{v}_0) - \boldsymbol{K}\boldsymbol{x}_a$, with $\boldsymbol{x}_a$ the background parameters vector, the formal retrieval equation follows this scheme:

$$\boldsymbol{x}\big(\gamma \boldsymbol{S}_a^{-1} + \boldsymbol{K}_T \boldsymbol{S}_y^{-1} \boldsymbol{K}\big) = \boldsymbol{y}\big(\boldsymbol{K}_T \boldsymbol{S}_y^{-1}\big) \qquad (4)$$

where $\boldsymbol{S}_a$ is the parameters' background covariance matrix, and $\boldsymbol{S}_y$ is the noise covariance matrix of the data. $\gamma$ is an additional regularization parameter *[Liuzzi et al., 2016; Carissimo et al., 2005]*, which acts as a tradeoff between the background values given to the parameters to be retrieved and the observations. Large values of $\gamma > 1$ will constrain the retrieval scheme more to the a-priori parameter values, As $\gamma$ approaches 0, the solution scheme tends to a constrained least-square. For $\gamma = 1$, the Rodgers' classical scheme is run.

OE provides a natural quantification of the uncertainty affecting the retrieved parameters, which we denote here with $\widetilde{\boldsymbol{v}}$, by computing the a-posteriori covariance matrix:

$$\boldsymbol{S}_{\widetilde{v}} = \boldsymbol{A}^{-1}\big(\gamma^2 \boldsymbol{S}_a^{-1} + \boldsymbol{K}_T \boldsymbol{S}_y^{-1} \boldsymbol{K}\big)\boldsymbol{A}^{-1}, \text{ with } \boldsymbol{A} = \big(\gamma \boldsymbol{S}_a^{-1} + \boldsymbol{K}_T \boldsymbol{S}_y^{-1} \boldsymbol{K}\big) \qquad (5)$$

and the Averaging Kernel (*AK*) of the derived parameters, which is defined as the sensitivity of the retrieved parameters with respect to their "real" value:

$$AK = \frac{\partial \widetilde{\boldsymbol{v}}}{\partial \boldsymbol{v}} = \big(\gamma \boldsymbol{S}_a^{-1} + \boldsymbol{K}_T \boldsymbol{S}_y^{-1} \boldsymbol{K}\big)^{-1} \boldsymbol{K}_T \boldsymbol{S}_y^{-1} \boldsymbol{K} \qquad (6)$$

The diagonal of the *AK* matrix contains values between 0 and 1, which indicate the sensitivity of the retrieval to each parameter. Low values could indicate either poor intrinsic sensitivity of the data themselves to the retrieval parameters, or too stringent constraints imposed to the variability of the retrieved parameters. The non-diagonal elements indicate the degree of correlation between parameters; ideally, these values are close to 0, or negligible with respect to the diagonal elements.

The preference of OE vs. e.g., a simpler regularized Least Squares approach lies in the fact that OE permits to specify definite constraints of the parameter space and weigh those against the data space. However, the adaptation of OE to this work is not trivial: the problem of retrieving 4 parameters from 5 data points is ill posed. To maximize confidence from retrieval, we have elaborated a procedure that dynamically adapts to the data properties. The algorithm is illustrated in Figure 4, and can be summarized as follows:

- By default, the parameter $\gamma$ is set to 5 to make convergence to the solution driven by parameters, and is increased anytime the data exhibit low information content or poor SNR. This choice serves to limit the influence of any bias affecting the data, and increases the convergence rate.
- To constrain the variables in a dynamic way, particle sizes are not retrieved when aerosol content is found to be too low, while covariances are tuned anytime the uncertainty (Eq. (5)) is beyond a certain threshold.
- Based on the known properties of vertical distribution of aerosols on Mars, we choose first guess values for particle sizes, concentration and variances variable with altitude. Values ranges are indicated in Table 1.
- Finally, independent of the occultation geometry (ingress or egress), the a-priori values of the 4 parameters for one spectrum are set to be equal to the retrieved values of the previous one, when converging for all four variables. Regardless, the vertical profile is the result of a sequence of single spectra retrievals.

Such progressive tuning is found to guarantee that ~85% of the analyzed data achieve convergence within the maximum number of iterations established in our scheme.

| Parameter | Variance | Initial guess value | Retrieval boundaries |
|---|---|---|---|
| Dust concentration $Q_{dust}$ | 5.0 ppm | 1.0 ppm | 0.0 – 100.0 ppm |
| Water ice concentration $Q_{ice}$ | 10.0 ppm | 1.0 ppm | 0.0 – 100.0 ppm |
| Dust particle radius $r_{dust}$ | 0.5 $\mu$m | h<20 km: 2.0 $\mu$m <br> h>20 km: 1.0 $\mu$m | 0.1 – 7.0 $\mu$m |
| Water ice particle radius $r_{ice}$ | 0.5 $\mu$m | h<20 km: 2.0 $\mu$m to 4.0 $\mu$m <br> h>20 km: 0.5 $\mu$m to 3.0 $\mu$m | 0.1 – 10.0 $\mu$m |

**Table 1**. *Summary of default first guesses and variances of the parameters as applied in the retrieval scheme. Water ice particle radius' first guess depends both on altitude and on the shape of the spectrum, which provides a rough estimation of particle size based on the characteristics of the absorption at ~3 $\mu$m. The unit ppm are given as mass units ([kg/kg]\*10^{-6}), and are based on the derived aerosol particle size, the implied extinction cross section for that particle size, and the measured extinction for each individual profile with altitude.*

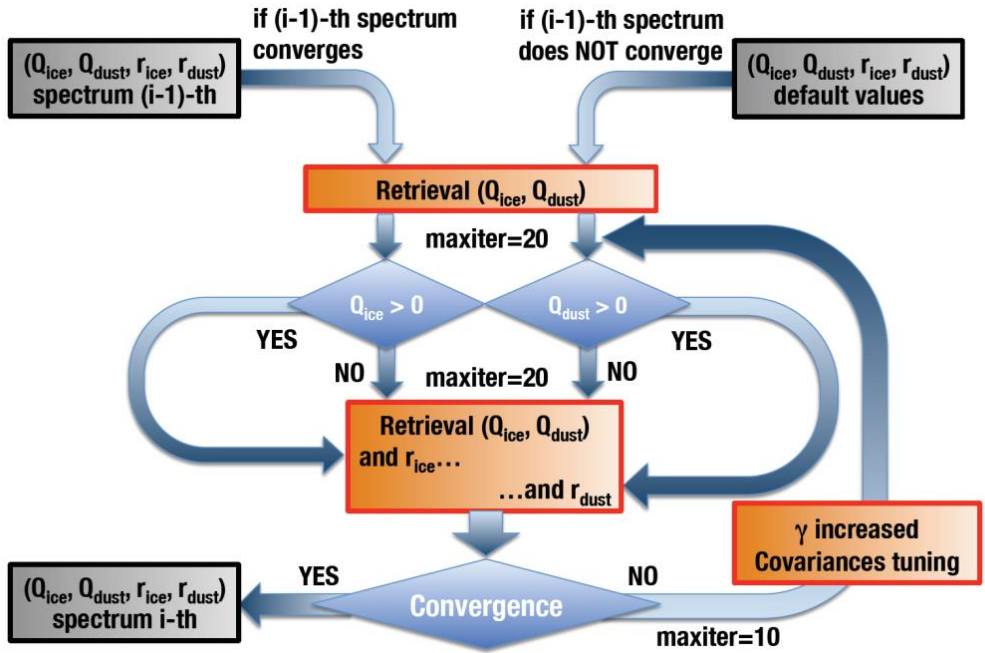

**Figure 4**. *Diagram explaining the aerosol retrieval algorithm. The diagram shows inputs and output of the retrieval applied to the i-th spectrum of an occultation. Orange boxes indicate actions taken by the retrieval algorithm, blue=verification of conditions, grey=input and output data.*

## 3 Dataset and results

The analyzed NOMAD dataset consists of all solar occultations acquired by the instrument in the period between the beginning of science operations (April 21$^{st}$ 2018) and April 2019 (L$_s$ ~162 of MY34 to L$_s$ ~15 of MY35), for a total of 1,781 (~172,000 data points) occultations. This dataset supports retrievals for the vertical distribution of aerosols from the lower atmosphere up to 110 km over a sparse set of geographic locations sampling the whole globe, expanding - especially in the upper atmosphere - the coverage of comprehensive studies made so far (e.g. *Clancy et al. [2019]*). The number of occultations performed by TGO is sufficient to track the overall temporal variability of aerosol distributions on Mars on a seasonal timescale. There are several cases of consecutive occultations at the same latitude in a restricted amount of time, which may enable the capability of tracking the temporal evolution of clouds (formation or dissipation) on timescales of days for a given local time.

To ensure the best compromise between spatial coverage and data quality, we have performed retrieval on all the data available, and filtered the results based on the following considerations:

- The optimal case for ice characterization occurs when 5 points of the broadband spectrum are uniformly distributed between 2700 and 4400 cm$^{-1}$, and when two of them fall in the range 3000-3400 cm$^{-1}$, which is key to accurately characterizing water ice particle size when it is lower than 1.5 μm. Therefore, we have considered only those results, which still comprise 83% of the total dataset.

- Aerosols can completely extinguish the solar radiation below 5 to 30 km, depending on the dust activity. In light of the discussion on the thermal-induced instrument systematics (sec. 2.2), transmittances can be heavily biased at low signal. Based on the quantification of AOTF-induced continuum variations, we have discarded all the spectra with a transmittance lower than 1%. Conversely, in the upper atmosphere, aerosol extinction becomes significant when the observed transmittance is lower than 99%.

### 3.1 Temporal trends

Figure 5 shows the temporal evolution of retrieved dust and water ice concentration vertical profiles. The two hemispheres are separated, and the orbit of TGO allows occultations only in specific locations, with slowly varying latitude as the relative inclination of the TGO orbit with respect to the surface change. To make the figures easier to interpret, we have reported the latitude of the occultations in the upper panels, with colors indicating the aerosol concentration averaged in two specific altitude ranges. Importantly, each data point in each panel combines the information from the retrievals by performing a weighted average of the nearby (in time and altitude) values, in order to present the data accurately. The two ranges correspond, conventionally, to the troposphere (10-40 km) and the mesosphere (40-80 km). The altitude of 40 km is typically where the Martian hygropause is located at perihelion, though its altitude varies, following the dynamics of expansion and contraction of the Martian atmosphere *[Slipski et al., 2018]*. We explicitly choose not to show such averages in the planetary boundary layer (up to 10 km) because of the lack of reliable data.

The first active process observed in the temporal sequences is the initiation of the June 2018 GDS (top panel). Figure 5 indicates that the storm spreads to all latitudes on a timescale of days, and dust abundances peaked around June 25th ($L_S$ 200) at the equator and Northern Hemisphere (NH). Our observations are generally consistent with the local characterization of the dust storm provided by Curiosity *[Guzewich et al., 2019]*, which presents a sudden increase of dust optical depth at $L_S$ 195. Based on the NOMAD retrievals, the lifted dust reaches a maximum altitude of 75 km in the equatorial region at $L_S$ 201, and a peak mass mixing ratio of 70 ± 10 ppm ($7 \times 10^{-5}$ kg/kg) at 45 km in the NH. High mixing ratio of dust is observed in the atmosphere after the initial outbreak of the GDS until the beginning of September ($L_S$ 250), with top of the dust layer slowly decreasing in altitude over time from 70 km ($L_S$ 203) to 40 km (perihelion) in the SH. The temporal sequence also shows the second dust event in January 2019 ($L_S$ 320 to 330, observed every MY; e.g. *Smith et al. [2009], Kass et al. [2016]*). The maximum altitude reached by dust in this case is 60 km in the SH, the peak mass mixing ratio of dust is 25 ppm at 40 km, and the dust abundance decreases more rapidly than during the GDS. Consistently with models, dust activity reduces greatly at the Equinox (Ls~0).

Dust abundances show a marked North-South asymmetry. A larger abundance of dust in the NH vs. SH has been noted in previous works *[Montabone et al., 2015; Smith, 2004]* in the same perihelion season (Ls 200-360), and by *Clancy [2003]* in contrast to the aphelion season (Ls 20-140). We note that this asymmetry is suppressed during the GDS, during which in the two hemispheres, on average, the same dust mixing ratio is retrieved below 40 km, with slightly larger abundances is the SH; no or little dust is observed, instead, in the lower atmosphere northern than 50 N. What emerges newly here is that asymmetry is present at high altitudes too (up to 60 km, black line in the upper inset of Figure 5 for dust) where the retrieved dust in the NH is, in average, much larger than in the SH.

The evolution of dust vertical distribution is reflected in the behavior of water ice and clouds (lower section of Figure 5). While the perihelion season ($L_S$ 180-300) is characterized by the constant presence of high-altitude water ice clouds at the terminator, these results show that the altitude at which they form is observed to abruptly change at the outbreak of the GDS. This is consistent with recent studies that predict (based on measured temperatures, *Heavens et al., [2018]*) or directly observe *[Vandaele et al., 2019, Fedorova et al., 2018]* sudden changes in the water vapor vertical distribution in response to heating produced by the dust lifted in the atmosphere.

This study demonstrates that the water ice cloud vertical distributions also respond promptly to the heating of the lower atmosphere due to the GDS, which leads to the displacement of water vapor saturation conditions to high altitudes. In a MY with a non-GDS perihelion season, water vapor saturation and ice cloud formation occurs over 30-50 km altitudes, such that water vapor abundance decreases rapidly into the mesosphere. During a GDS (as in MY34), elevated dust leads to increased temperatures to well above 40 km, and so pushes the hygropause trapping of water vapor to higher altitudes where water vapor remains unsaturated due to increased saturation pressure *[Neary et al., 2019]*. Water vapor is then transported into the mesosphere, where colder temperatures lead to widespread, optically thick mesospheric water ice clouds. NOMAD water ice profiles demonstrate the process in detail: in the initial phase of the GDS ($L_S$ 195-225), water ice clouds are not observed to be present below 40 km altitudes, other than in thin hazes. Such strong vertical transport is also summarized by the two top right insets of Figure 5, where it is shown the comparison between water ice concentration during the storm (black line) and right after that (grey patch).

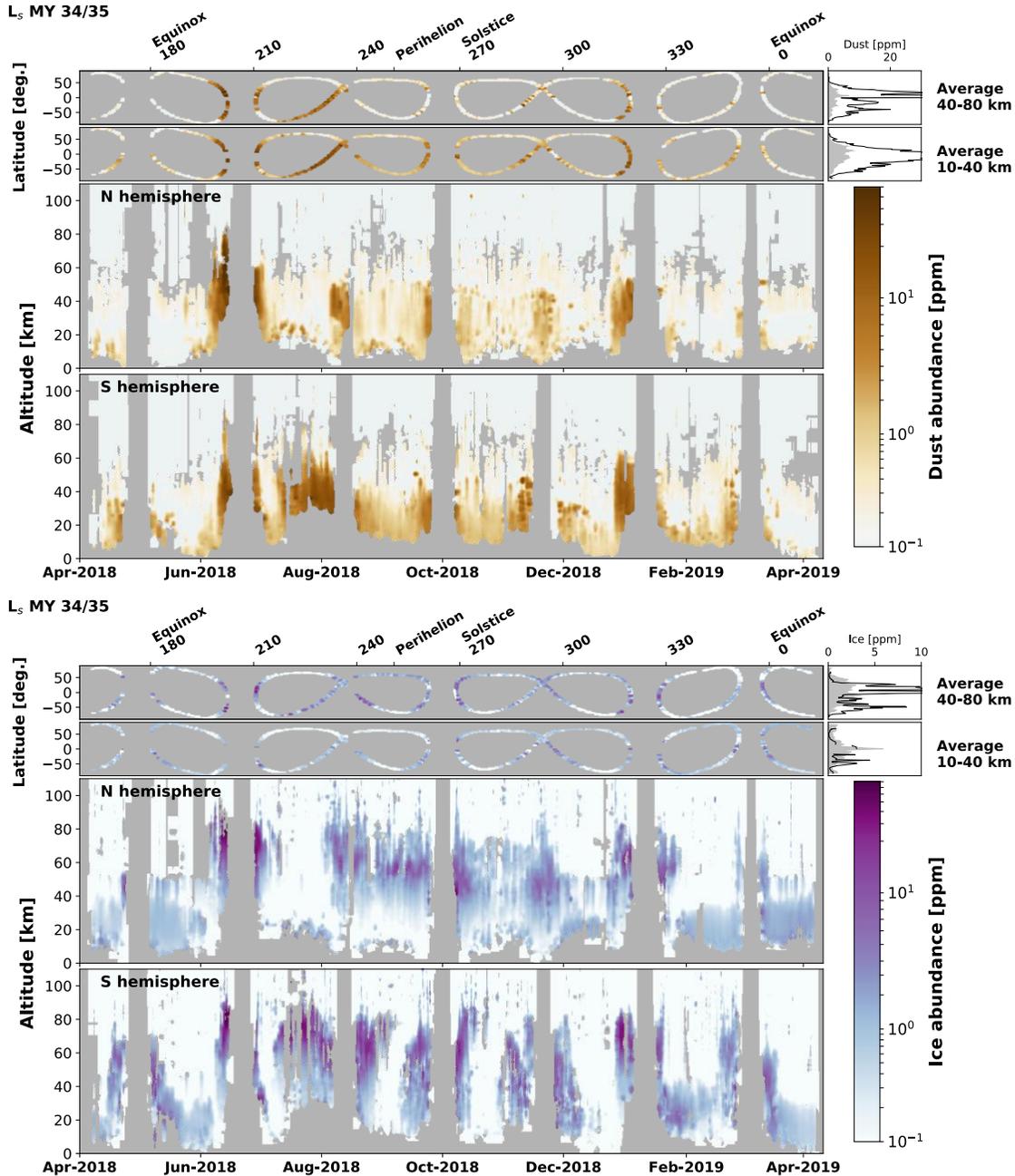

**Figure 5**. *Retrieved dust (top) and water ice (bottom) concentration (in parts per million of airmass) vertical profiles and their evolution with time. All panels show a weighted average of the retrieved values from the surface up to 110 km, divided by hemispheres. Grey areas correspond to no data or excessive opacity (poor SNR). The two top panels report, for dust and ice, and with the same color scale, their averages in latitude in two altitude ranges, to highlight the actual latitude where occultations occur. The two insets on the top right represent the grand average (over time, on the two altitude ranges) of concentrations during the 2018 GDS (June 10th to August 15th, black line) and out of the 2018 GDS temporal range (grey fill).*

Over the range of observed latitudes (Figure 5), the average water ice condensation altitude is found to slowly decline as the dust storm dissipates, from a maximum of ~90 km ($L_S$~200) to ~50 km ($L_S$~255), following the lowering of the top of the dust layer (Figure 5, top). By $L_S$ 290 the altitude at which clouds form returns to pre-storm values (45 km). This observation confirms what emerges from previous model simulations (e.g. *Kahre et al. [2008]*) and earlier observations, that the atmosphere takes several tens of sols to relax to a pre-storm thermal condition. In this case, NOMAD observations suggest that such process takes an interval of ~80° $L_S$ (i.e., 150 sols). Similar dust and ice behaviors are also present in the 2019, $L_S$~320 storm, where the water condensation altitude raises by 20 km compared to pre-storm values. In this case, however, dust opacity declines more rapidly, as does the cloud formation altitude.

As discussed in Section 2.2, measurement and retrieval uncertainties limit information about the aerosol type (dust vs. ice) in the lower atmosphere during the GDS. While it is fair to assume that the retrieved opacity is primarily due to dust, it is still not possible to rule out the presence of small amounts of water ice between 10 and 30 km during the most intense phase of the dust storm. Hints of low altitude cloud formation are visible in the NH panel around $L_S$ 225 and perihelion; they may form by cooling of the lower atmosphere by decrease of solar flux at the surface in the VIS and IR, and by saturation of water vapor. Both aspects are documented by Curiosity observations *[Guzewich et al., 2019]*, which detected a 40 K air temperature drop at the beginning of the storm at surface level. Nevertheless, NOMAD data do not imply any clear detection of water ice between $L_S$ 215 and 240 at any altitude below 30 km. In the northern mid-latitude region (winter), clouds start to re-appear at $L_S$ 270 between 10 and 30 km, and at $L_S$ 245 in the near-equatorial region of the SH.

The vertical and seasonal evolution of aerosol particle sizes derived from NOMAD solar occultations, in particular during the GDS and its dissipation, provide further insights into aerosol microphysics associated with Mars GDSs. Figure 6 shows retrieved particle size for water ice (top) and dust (bottom). It is important to stress that both plots include only those retrievals corresponding to cases where ice and dust are consistently separated, and where water ice and/or dust abundance is high enough (i.e. produces extinction above the level of the systematics of the instrument) that the particle size is meaningful.

Water ice particle sizes display a known gradient in vertical distribution within the 30-50 km atmospheric region *[Guzewich et al., 2014, 2019]* and into the mesosphere *[Clancy et al., 2019]*. NOMAD retrievals indicate mesospheric clouds are characterized by small particle sizes, mostly between 0.1 and 0.5 µm in the core of the clouds, with sizes decreasing as altitude increases (as spatially and seasonally mapped in *Clancy et al. [2019]*). An important feature found in water ice size retrievals here is a pronounced asymmetry between NH and SH below 50 km, after the most intense phase of the GDS ($L_S$ 240 to 300). During this time, the likely decreasing temperatures in the middle atmosphere enable greater water ice abundances and nucleation around the residual suspended dust (more abundant in the SH than the NH), forming bigger water ice particles, with sizes up to ~1.5 µm at the base of mesospheric clouds.

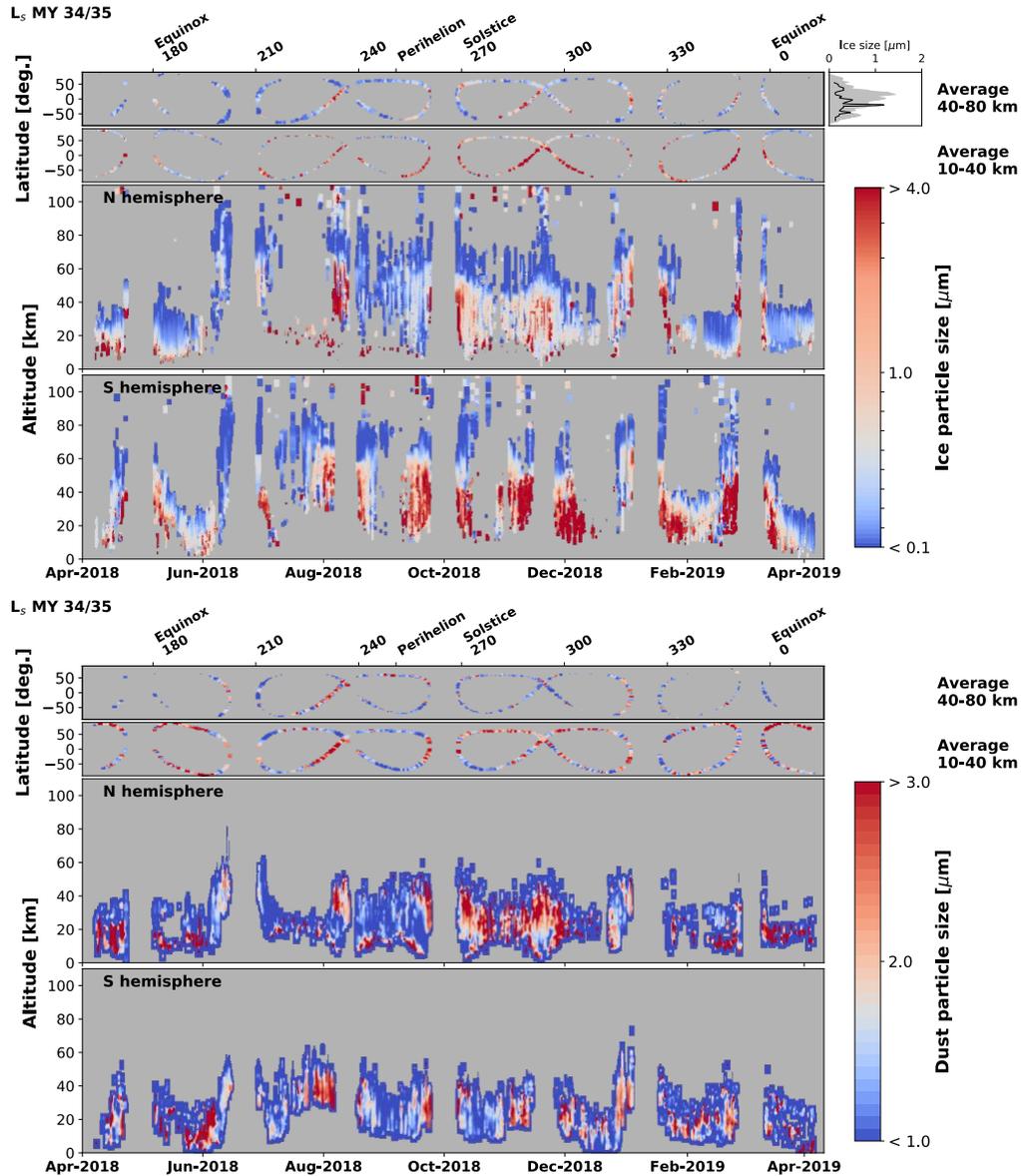

**Figure 6**. *Same as figure 5, but for water ice particle radius (top) and dust particle radius (bottom). Only significant top right plots are reported, since during the GDS no or very few low altitude clouds are found.*

Water ice particle radii in the lower atmosphere are poorly characterized due to particle radii correlations resulting from NOMAD dust and ice retrievals. Derived particle sizes above 30 km, which are less influenced by dust contamination, exhibit sizes between 1.0 and 2.0 µm (roughly consistent with *Guzewich and Smith [2019]*). After L$_s$ 260, water ice particle sizes in the lower atmosphere show a strong dependence on latitude, with sizes above 2.0 µm in the SH (warmer and more dusty), and smaller than 1.0 μm at latitudes northward of 40 N. After the GDS, particles larger than 1.5 μm are confined only between latitudes of 50 S and 50 N.

In the perihelion period, the retrieved latitude distribution of water ice particle size agrees with previous studies *[Guzewich & Smith, 2019]*, where the NOMAD retrievals extend the ice particle size climatology in coverage and altitude. Single, collocated observations with ACS *[Vandaele et al., 2019, Luginin et al., 2019]* show also a high level of consistency with these retrievals. In particular, water ice retrievals are highly consistent both in pre-GDS observations in the NH (with a steep gradient in the water ice particle size, from 3.5 to <0.5 μm), and during the GDS in the SH. The agreement between retrieved dust particle sizes is instead more challenging to evaluate, since the two instruments cover different wavelengths, hence their sensitivity to dust particle size is not the same.

An interesting feature seen in the temporal series of water ice particle size is located right after the onset of the GDS, with large water ice particles (~3 μm) between the equatorial region and mid-N latitudes, at altitudes between 25 and 60 km, at Ls~230 (see for comparison bottom of Figure 5). In the same time frame, we retrieve large dust abundances (top of Figure 5) with particle sizes larger than usual (>2.0 μm, Figure 6) between 25 and 50 km at the same latitudes. This resembles, in location and altitude, what happens during local highly convective dust storms *[Spiga et al., 2013]*, in which efficient transport of dust at altitudes of 30 to 50 km occurs. Despite the lack of specific information about atmospheric temperature, it can be seen that this suspended layer of dust persists for several Martian days before declining, another typical behavior of rocket dust storms. This finding challenges the work in *Spiga et al., 2013*, that indicate rocket dust storms are less likely to occur in the dusty, warm perihelion season as static stability maximizes with elevated dust globally, but agrees with more recent findings during MY34 GDS *[Heavens et al., 2019]*. Other results *[Wolff and Clancy, 2003]* obtained with TES observations show such variable increases in dust particle sizes during the MY25 GDS, likely in association with intense regional dust lifting. The rest of the values we retrieve for dust particle size is found to be in good agreement with the literature (1.5 μm, e.g. *Clancy et al. [2003]; Wolff and Clancy [2003]; Smith [2004]; Guzewich et al. [2014]*); however, apart from the cases in which large dust abundances are observed, NOMAD retrievals of dust particle size are generally less accurate than the water ice ones.

## 3.2 Uncertainties and biases

The factors that can limit the reliability of the retrievals are mostly related to signal intensity, the degree of correlation between the retrieved variables, and the assumptions about the vertical structure of the atmosphere. The effects of low signal intensity are mitigated by the strategies explained above, including not performing any retrieval on spectra whose transmittance is below 1%, which is the order of magnitude of the temperature-induced variations in the single-order spectral continuum. Such variations are typically monotonic with time, and because the orders are simultaneously measured, they impact all orders in the same way. The retrieved values are not dramatically biased when the flux measured by the instrument is above 1%. However, below that threshold, the spectral features are significantly distorted, and dust and water ice extinctions cannot be separated effectively. The values retrieved in the middle and upper atmosphere are highly reliable, since the spectral contrast is much larger than any temperature-induced artifact.

The derived uncertainties for each parameter are showed in Figure 7, with NH and SH merged. It is apparent that water ice is better characterized than dust. Mesospheric and polar ice abundance is derived with accuracy between 10%-25%, that degrades to 40%-50% for the majority of the values below 30 km. This is reflected in the accuracy of water ice particle size (panel (c)), which is around 0.05 μm for mesospheric ice, around 0.2 μm for the polar regions, and 0.7 μm below 30 km. This suggests that, for mesospheric ice clouds, the retrievals are able to consistently discriminate particle sizes whenever the differences are 0.1 μm or bigger (0.1 from 0.2 μm, 0.2 from 0.3 μm, etc.).

Figures 7(b) and (d) highlight the issues in retrieving dust parameters: dust abundance accuracy is between 10% and 20% during the dust storms, but degrades rapidly above 40 km, indicating confidence in extracting the dust abundance in the lower atmosphere, separate from water ice. Similarly, dust particle size can be retrieved only in cases in which dust optical depth is high.

Independence between the uncertainties of water ice and dust is a good indicator that, with the constraints imposed on data selection, we are able to separate dust and water ice on the remaining data. This is reflected in the AKs of Eq. ( 6 ) produced with each individual retrieval (panels (e) to (h)). Water ice abundance is always retrieved consistently and independently of the other parameters; hence the diagonal element of the AK corresponding to water ice concentration is always close to 1 (Figure 7(e)), and the corresponding non-diagonal elements are always close to 0. Water ice particle size are also robustly retrieved for small particles (AK~1), while in general AK<0.9 for size >1.5 μm. The off-diagonal elements between size and abundance can be larger than a few percent, indicating that a non-negligible portion of the information about particle size comes, actually, from the concentration. This effect becomes more severe as the water ice extinction decreases.

The analysis of AKs confirms the issues in obtaining independent information about dust abundance and size (Figure 7 (f) and (h)): while the abundance diagonal term of the AK is often close to 1 (and decreases with the abundance), the particle size term is >0.75 only for large concentrations. Moreover, the absolute values of off-diagonal terms are often >0.1 for both quantities, indicating a high degree of contamination between the two. This is expected, based on the considerations reported in Section 2.3.

The last point of systematic uncertainty is related to the assumption about the a priori atmospheric state. We assume gases abundances and temperatures directly from the GEM model *[Neary et al., 2019]*, that reproduces the MY34 GDS scenario. Indeed, during large dust storms, the heating of atmosphere produced by lifted dust alters the scale height significantly, and with that the $CO_2$ column densities. This does not directly impact the accuracy of the retrievals, which are strongly opacity-based, but the abundances we report are referred to the atmospheric total density, a fact that automatically may introduce a bias, should the actual pressure differ from GEM. Nevertheless, this does not change the general behavior of the retrievals (clouds and dust layers altitude, temporal trends, etc.), and impacts only those analyses that quantitatively compare gases and aerosols, which go beyond this work.

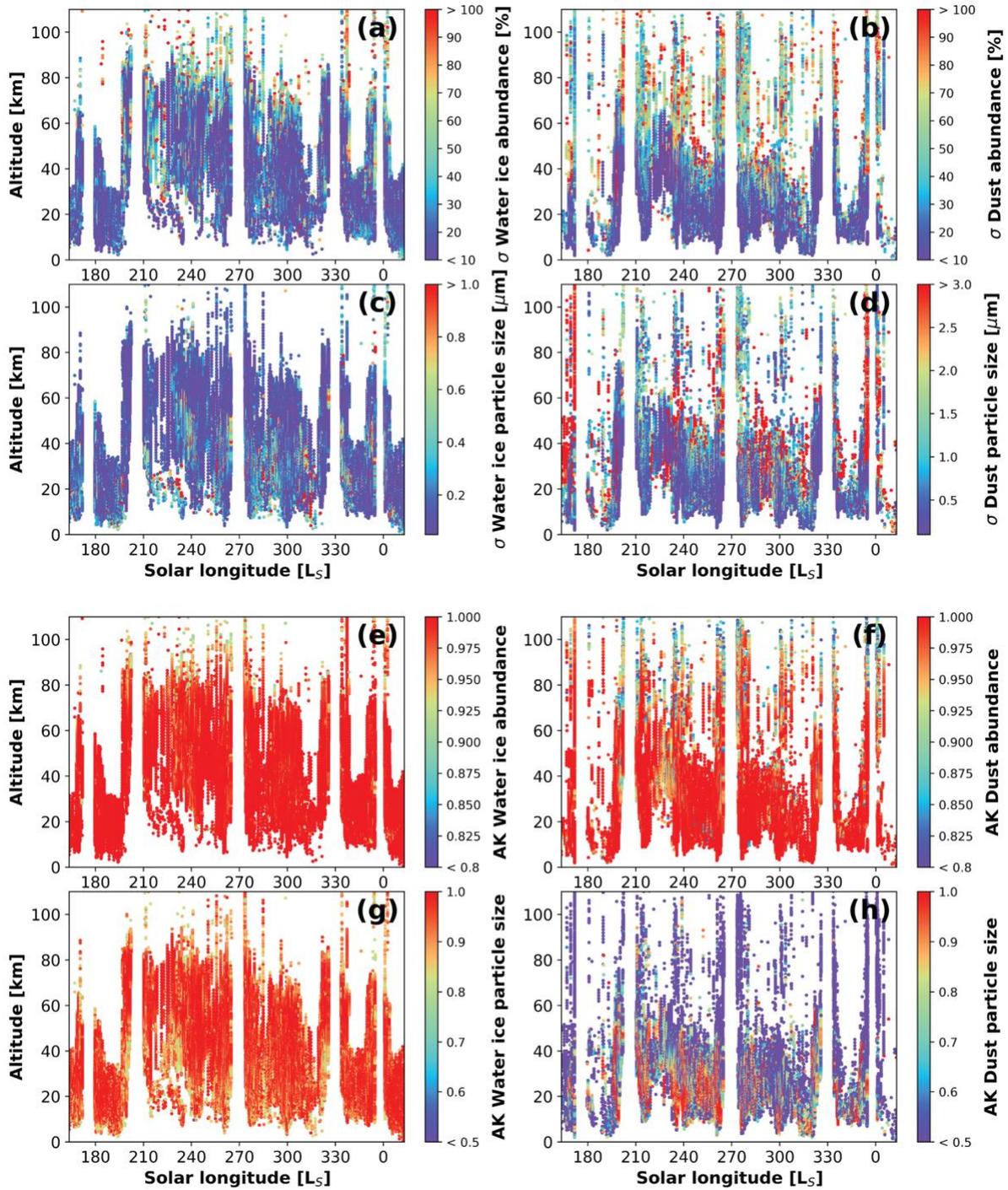

**Figure 7**. *Uncertainties (a-d) and AKs (e-h) associated to each of the retrieved parameters. Uncertainties for water ice abundance (a) and dust abundance (b) are reported as relative accuracy, those for water ice particle size (c) and dust particle size (d) in microns.*

**3.3 Diurnal results and latitude distribution**

The major strength of this dataset is associated with the unique profiling sensitivity associated with the very high opacities typical of a GDS, thanks to the simultaneous retrieval of concentration and particle size, as demonstrated by the results and accuracies reported so far. However, in general NOMAD alternates occultation pairs each orbit, one taken in each the SH and NH, one at dawn (local time 6hrs) and one at dusk (18hrs). This enhances the scientific values of these retrievals, since they complement sun-synchronous orbiter observations performed to date (MGS and MRO), which were acquired at 2 to 3 am/pm in both Nadir (column values) and Limb geometries (vertical profiles). Thus NOMAD vertical profiles can reveal new information about the aerosol dynamics and microphysics in two local times.

The dawn and dusk results are presented in a series of plots in Figure 8, where the retrievals of water ice and dust mixing ratio have been organized in bins of 30 Ls, the minimum temporal interval consistent with substantial latitudinal coverage. For those times where intense dust activity is retrieved, we also reported the thermal tides simulated according to MCD v5.3 database *[Millour et al., 2015]*, for a dust scenario corresponding to MY25, in which a GDS similar to the one in MY34 occurred. Thermal tides are defined according to *Lee et al., [2009]* as the zonal mean of the difference between temperatures at 6hrs and 18hrs. Wherever such differences are large, we expect enhanced cloud formation at dawn/dusk with respect to dusk/dawn. This analysis can be used as a proxy to evaluate if the retrieved water ice clouds are consistent with the expected thermal behavior of the atmosphere during a GDS.

The GDS featured in Figure 8(b) confirms the conclusions already drawn from the analysis of the temporal trends, showing that cloud formation altitude raises globally. In this phase, aerosol concentration is inferred only at altitudes above 30 km in the equatorial region, since the opacity in the lower atmosphere causes the signal to drop below 1%. During the most intense phase of the GDS, the vertical structure of water ice clouds is surprisingly similar between dawn and dusk. At equatorial and mid-latitudes, water ice haze form at 30 km by nucleation around lifted dust particles, while the clouds bulk form between 65 to 90 km. There is a strong asymmetry in the latitudinal distribution of clouds between dawn and dusk (panel (b)) where no high-altitude clouds are detected poleward of 40° S and 50° N at dawn, while high-altitude clouds extend up to 75° S at dusk. This is an indirect demonstration of a strong enhancement of the downwelling branch of the meridional circulation during the GDS, which increases the saturation pressure at the hygropause, reinforcing the transport of water vapor towards high latitudes (e.g. *Neary et al. [2019]*), and favoring mesospheric clouds formation in the late day. Enhanced cloud formation at high S latitudes at dusk, as well as some of the high altitude clouds seen at 60 N at dawn, are also consistent with the expected diurnal thermal tides.

A similar enhancement in meridional circulation can be seen in panel (e), which corresponds to the outbreak of the second storm (January 2019). In this case, there is also significant asymmetry in the NH vs. SH distribution of dust, which is notably less abundant in the NH than in the SH. Based on the available retrievals, it does not look like water ice cloud formation is consistent with the expected thermal tides. This offers a hint to characterizing differences in the cloud formation process at various latitudes, and the role of thermal tides compared to other condensation mechanisms that account for heterogeneous nuclei sources (*Plane et al., [2018], Hartwick et al., [2019]*).

Immediately after the most intense phase of the storm (panels (c) and (d)), the enhanced meridional circulation starts to vanish, and the distribution of clouds at dawn and dusk is still slightly asymmetric (c) at southern latitudes, returning to symmetric at $L_S$ 270 (d). The few exceptions to this symmetry are consistent with diurnal thermal tides, which appear to be stronger at N high latitudes at dawn and S high latitudes at dusk. The thermal effects due to dust are still visible, as the clouds that appear at dawn (presumably forming at nighttime) are thicker than the ones on the dusk side (panel (d)), where cloud formation is generally less efficient, especially at the equator above 50 km. Despite this, between the two storms NOMAD reveals the presence of a frequent dawn-dusk side terminator cloud belt, which extends from 45° S to 60° N at dawn, and from 60° S to 60° N at dusk (panel (d)).

Panels (c) and (d) reveal a strong asymmetry in the distribution of dust between dawn and dusk, particularly in the declining phase of the GDS where ice clouds are frequently present. This asymmetry was detected in previous studies (e.g. *Guzewich et al. [2013]* using TES limb data), but given the particle size and precipitation timescales, it was assumed to be unlikely that deposition processes occurred on a diurnal timescale. Instead, this asymmetry is likely due to nighttime scavenging of water ice on dust particles, rather than daily variation of the dust vertical profiles. Indeed, retrievals in panel (c) suggest that water ice nucleates at nighttime between 30 and 50 km, coating dust particles. During the day, solar radiation enhances dust-induced heating, which causes ice to sublimate between 30 and 50 km, where the dust mixing ratio is maximal (panel (c), right) As a result, ice is observed mostly in form of thinner haze at dusk, and its sublimation makes dust nuclei observable again. This is compatible with dust-induced heating in the same atmospheric region, which is likely to occur as long as significant amounts of dust are in the middle atmosphere.

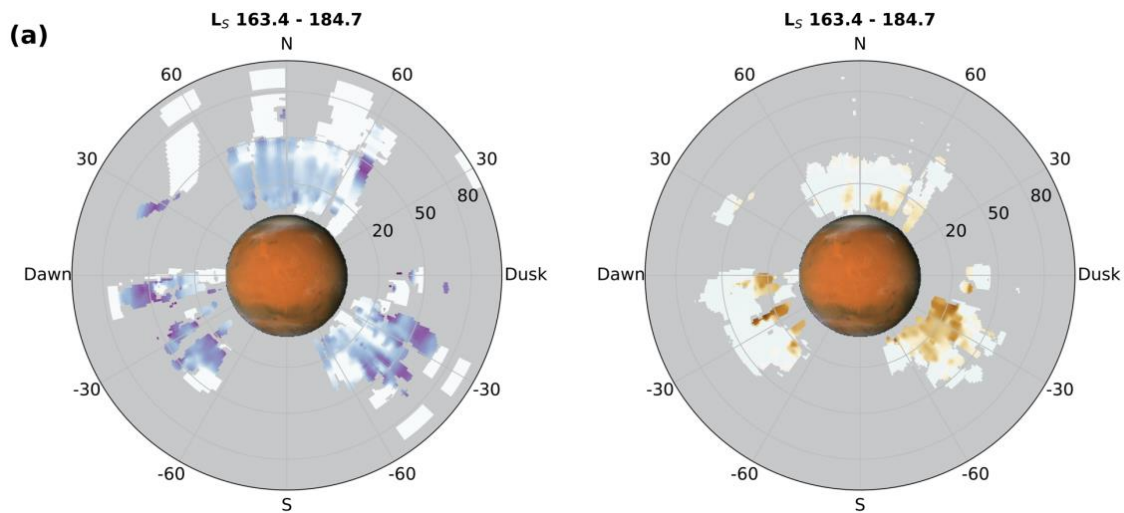

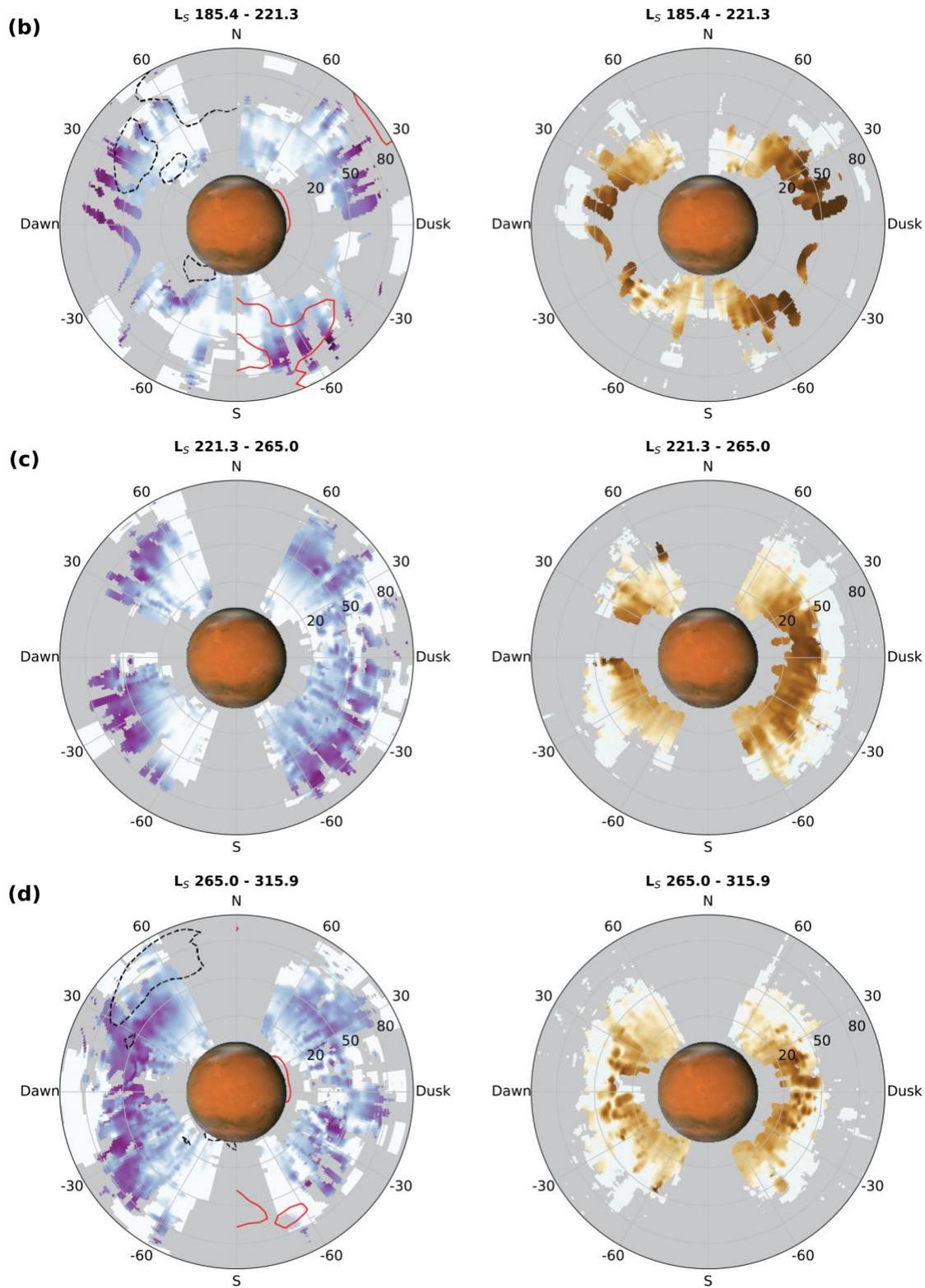

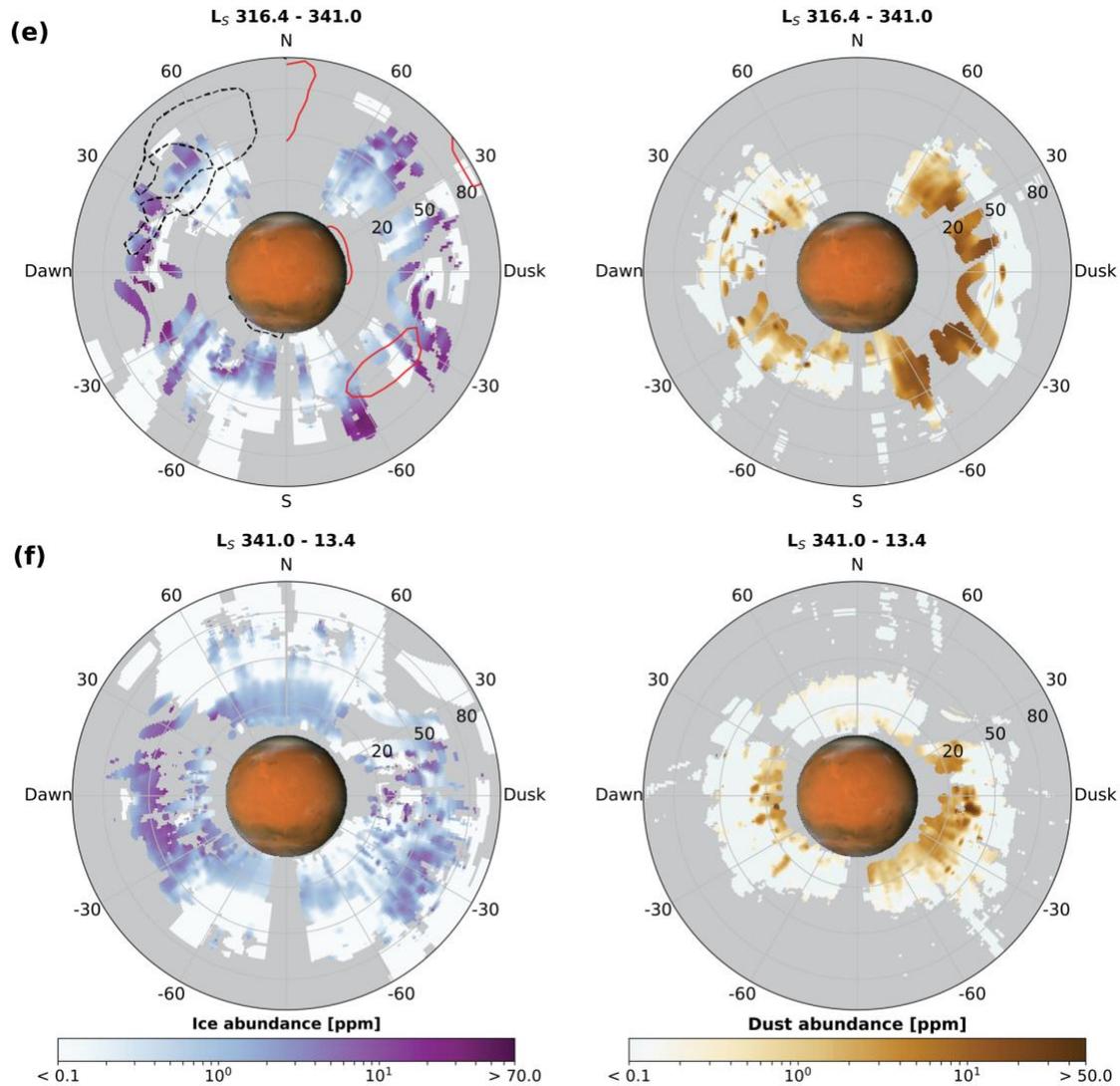

**Figure 8**. *Latitude distribution of water ice (left) and dust (right) abundances in time intervals of $L_S$ 30 for dawn (plotted on the left side of the planet) and dusk (right side). The division is useful to highlight the impact of the GDS and the Jan 2019 dust storm. Gray areas indicate no data or poor SNR. The radius is the altitude above the surface [km], and goes from 0 to 100 km. For panels (b), (d), and (e) we plot the contours of areas where strong (>4 K) thermal daily tides are expected, based on MCD simulations for MY25.*

In the context of cloud formation processes, more interesting elements emerge from the comparison between the GDS and the January 2019 dust event. While there are numerous similarities between the spatial distributions of the mesospheric clouds that form during the peak of the two events, the average particle sizes are different (Figure 9). Limiting our comparison to the altitude range in which clouds form in both cases (40 to 80 km), we note that at all latitudes particle sizes are different, with values in the range 0.1-0.7 µm for the 2018 GDS and 0.1-2.0 µm in the January 2019 event. This significant difference is attributable to the differing water vapor abundance in the upper atmosphere during the two events. Indeed, because of the season and lower dust activity of the January 2019 with respect to the 2018 GDS, we effectively observe a lower water vapor abundance in the mesosphere during the January 2019 event, compared to the 2018 GDS, when water was observed at 80 km *[Vandaele et al., 2019]*. The water profiles shown in Figure 9 are retrieved from NOMAD full resolution data, combining the retrievals from different diffraction orders *[Aoki et al., 2019]*, and are the global averages of the profiles during the onset of the two storms. These phenomena are supported by *Hartwick et al. [2019]* that discuss the relation between the abundance of water vapor in the upper atmosphere and the availability of condensation nuclei. In this case, the 2019 dust event is characterized by less water vapor in the upper atmosphere than the GDS. Moreover, Figure 8 indicates that much less dust is available in the upper atmosphere in the 2019 event than during the GDS. The combination of these two leads to less competition in the condensation process, yielding to formation of larger ice particles (> 1.0 μm), as observed during the 2019 dust event.

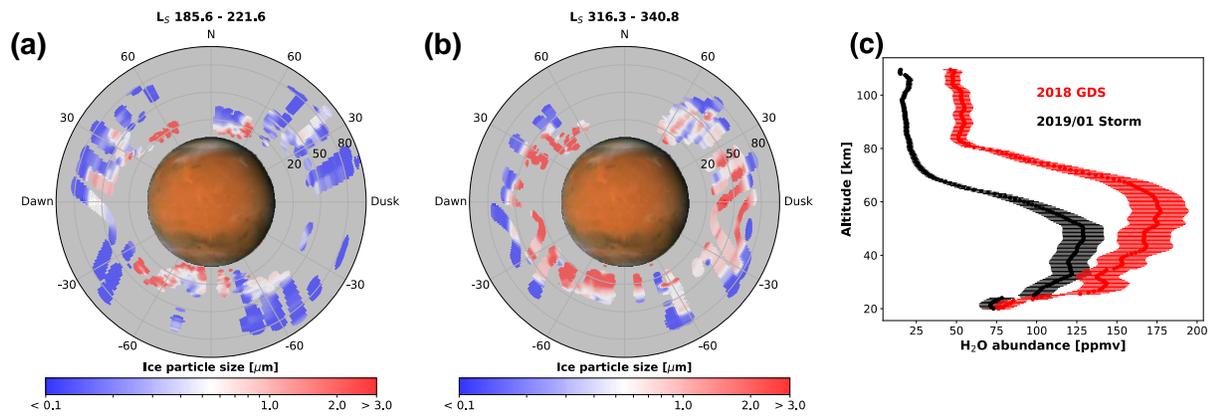

**Figure 9**. *In the same format as* **Error! Reference source not found.** *we illustrate water ice particle sizes retrieved during the GDS (left) and January 2019 (center) dust storm. On the right, the global averages of the water vapor vertical profiles retrieved from NOMAD data for the 2018 GDS (red) and the January 2019 storm (black).*

## 4 Discussion

### 4.1 Spatial and temporal distribution of clouds

The results illustrated herein contribute to the discussion of atmospheric circulation and formation mechanisms for water ice clouds on Mars. Although a detailed comparison between our results and global circulation modeling is beyond the scope of this paper, we will make broad comparisons with the existing literature on these topics. The latitudinal and temporal distribution of water ice clouds during the 2018 GDS (Figure 8), suggests a correlation between the enhancement of the water vapor circulation towards Southern high latitudes during the GDS (seen in *Neary et al. [2019]*) and the formation of high water ice clouds at high latitudes. Despite this, our results indicate that there are no significant differences in the ways that clouds are distributed at high latitudes in a global dust storm and non-GDS situation. The ideal reference for comparison are MCS retrievals *[McCleese et al. 2010]* during MY29. In both cases (Figures 16 and 17 of *McCleese et al. [2010]*), there is a clear break in the formation of high-altitude clouds in the NH, which occurs between mid-latitudes (45 to 60 N, Figure 8 (b) and (c)) and in the polar region. Clouds are seen only below 40 km both in MCS non-GDS retrievals and in this analysis.

Such comparisons show several differences for high-altitude water ice clouds. MCS results in the SH summer report that the top and the bottom level of most clouds are roughly separated by one order of magnitude in pressure, which corresponds to more than two scale heights (25 to 30 km). As seen here though, the altitude distribution is frequently much more complex (Figure 5 and Figure 8(c)), with clouds at the tropics and mid southern latitudes that extend for more than 40 km. The vertical extension is also subjected to dawn/dusk variations, which in the presence of a GDS are likely to occur, as discussed previously and suggested in other works (e.g. *Smith [2009]*). This depicts a complex cloud formation mechanism, which in the perihelion season considers both nucleation around dust and vertical and diurnal temperature gradients from the boundary layer through the mesosphere.

Comparison with MCS non-GDS retrievals also confirms that the rapid increase of cloud formation altitudes, seen in both the dust storms observed with NOMAD, is a storm-induced effect. Such a dramatic increase is not seen in MCS retrievals, as there is no tangible sign of a sudden variation of the vertical distribution of the clouds in the different time slots covered by MCS (L$_S$ 180-225-270-315, Figures 16 and 17 of *McCleese et al. [2010]*). Low-altitude clouds presented in this work are in substantial agreement with MCS retrievals between L$_S$ 180 and 360. Both retrievals (GDS and non-GDS), cannot identify significant water ice extinction below 25 km around L$_S$ 225, at nighttime for MCS and on the dawn terminator for NOMAD. The same agreement is found between daytime MCS retrievals, and NOMAD at dusk, with significant differences only in the mid-southern latitude towards L$_S$ 270, where MCS found no significant amounts of water ice below 50 km. Such differences can also be explained by the enhancement of the downwelling branch of the meridional circulation during the GDS. No discrepancies are found between MCS and NOMAD retrievals at the Equinox (L$_S$ 0), where clouds form at decreasing altitudes as latitude increases. In this case, both retrievals show that at low latitudes water ice forms as low as 15 km at nighttime (dawn), while during the day the condensation altitude increases because of the solar heating, and clouds form only at 30 km.

## 4.2 Water ice particle size: a comparison with CRISM retrievals

A great deal of information about the particle sizes of water ice in the atmosphere of Mars has been obtained by analyzing the limb data of the CRISM spectrometer. The present work increases the data coverage both in time and space with respect to CRISM, providing a more comprehensive assessment and validation of water ice particle sizes and their vertical and temporal variations.

The vertical structure found herein is consistent overall with the results presented in *Clancy et al. [2019]*. Mesospheric water ice clouds exhibit a narrow range of particle sizes (0.1 to 0.3 µm), however we have a larger number of cases where the retrieved average particle size is 0.1 µm (lower boundary imposed in the retrieval), as NOMAD retrievals are indicative of a particle size <0.1 µm. In the work of *Clancy et al.*, there is a distinct decrease in detections for particles smaller than 0.1 µm, because smaller aerosols are difficult to discriminate against bigger particles in CRISM data, and can only appear as a continuum scattering component. While our results agree with the general conclusion that particle sizes decrease with increasing altitude, we detect a significant number of water ice clouds in the mesosphere, where the vertical structure of particle sizes is more complex. In such cases, particle sizes exhibit local maxima at the center of the cloud layer, and then decline rapidly towards the cloud top (Figure 9(a)). While it is difficult to track the spatial distribution of such cases, they appear more frequent in optically thick clouds characterized by the presence of one main layer, which form above 60 km. In any case, the water ice clouds presented in the CRISM study are typically discrete layers, presumably formed in the cold phase of gravity or tidal waves *[Clancy et al. 2019]*, while NOMAD is frequently observing extended cloud hazes during the dust storm, which are situated below the bulk of mesospheric clouds.

Similar differences in complexity are seen in comparison with previous works *[Guzewich et al., 2014; Guzewich & Smith, 2019]* regarding water ice clouds in the lower atmosphere. In particular, Figure 6 (top) informs variations of water ice particle size in two different altitude ranges. Our lower atmospheric retrievals (10-40 km) show substantial disagreement with CRISM retrievals. Considering only those cases where retrieved water ice concentrations are significant, the average retrieved particle size (30 S - 30 N, 20-40 km) is 2.7 $\mu$m at perihelion (L$_S$ 240-260) and 2.2 $\mu$m at the equinoxes (L$_S$ 170-190 and 350-10, MY 35), in contrast to values respectively of 2.1 and 1.7 $\mu$m from CRISM. While the perihelion difference can be attributed to the dynamics related to the GDS, the discrepancy observed during the Equinoxes is more difficult to explain, although it seems consistent with the sharp vertical gradients we see in particle sizes. Vertical variation of particle size appears to be much steeper than CRISM retrievals *[Guzewich et al., 2014]*. On one hand, limb observations show greater sensitivity to characteristics otherwise difficult to constrain (i.e., particle phase function, shape, etc.) so the difference between retrievals might not be really significant. On the other hand, there is the possibility that water saturation is sufficiently high or low that particle growth is more or less efficient than expected. This aspect deserves to be further investigated by dedicated modeling work.

In general, the frequent detection of mesospheric water ice clouds composed of small particles poses some interesting questions related to their impact on the radiative balance in the upper atmosphere, and the lower altitude layers. Previous works (e.g. *Madeleine et al. [2011]*) have already shown the presence of a permanent cold bias around 0.1 mbar, which can be mainly attributed to the exclusion of the radiative effects of high altitude water ice clouds composed by small particles, especially during the perihelion. In fact, this population of clouds cannot be captured by the unimodal size distribution assumed in dust transport models included in some GCMs. This modeling limitation has been questioned recently in *Hartwick et al., [2019]*, where it is shown how a model including meteoric smoke can account for such observations. The retrievals we have presented here do not provide a final word on nucleation processes in the upper atmosphere, since the thermal information is not fully integrated into this analysis. However, the dawn/dusk NOMAD measurements can certainly inform on those processes in a complementary fashion to the MCS 3 am/3 pm retrievals, which have been used as a benchmark in the work by *Hartwick et al*. In addition, the capability of NOMAD observations to extensively constrain water ice and dust abundance and properties up to 100 km is of great importance in this context. The present work is only the most recent showing the persistent presence of such clouds at many latitudes.

## 5 Conclusions

NOMAD measurements contain a breadth of information on Martian atmospheric aerosols. By using all the available data taken by NOMAD in Solar Occultation, we have retrieved vertical profiles of water ice, dust and their particle sizes, with a resolution around 1 km, a maximum vertical sampling of 600 m, from the lower atmosphere to 110 km. To accomplish this, we have developed a robust retrieval methodology to treat NOMAD SO broadband data. Given the number of available points, this is an under-constrained problem, the resolution of which constitutes an important piece of work. We have generally obtained robust results, characterized the errors and the information content, highlighting the consistency of retrieved water ice properties, and the caveats associated to the dust retrievals. In general these results indicate that when the observed transmittance is above 1%, retrievals are robust enough to separate water ice from dust, and quantify their microphysical properties.

We have analyzed NOMAD data from April 2018 to April 2019, for a total of 1,781 profiles. This period encompasses the 2018 GDS, which was observed to have tremendous effects not only on the vertical distribution of water vapor, as shown in previous studies, but also on water ice cloud formation. The rapid lifting of the water ice condensation altitude is a peculiarity of dust storms, since it has not been observed in a non-GDS situation previously. This effect has been seen to last for a long period (80° Ls) after the onset of the storm. Clouds are observed as high as 90 km at the beginning of the GDS, while dust elevates up to 70 km.

Water ice clouds have been observed at dawn and dusk. There are remarkable differences between the two, with optically thicker (larger concentrations) mesospheric clouds at dawn than dusk due to nighttime condensation of water vapor. The combination of dust and ice observations between dawn and dusk reveals how dust grains are subjected to nighttime scavenging by water ice. Observations suggest that this process decreases in intensity as the GDS dissipates. Dawn vs. dusk analysis also reveals the presence of dusk high latitude mesospheric clouds in the SH during the most intense phase of the GDS, which is compatible with a strong enhancement of the downwelling branch of the meridional circulation.

We characterized the particle sizes of mesospheric water ice clouds with a precision around 0.1 μm. The majority of water ice clouds particle size vertical profiles exhibit sharp vertical gradients. Specifically, mesospheric water ice particles have sizes between 0.1 and 0.5 μm, which decrease with altitude. However, the comparison with literature shows previously undetected complexities in the vertical profiles of water ice and particle sizes, which constitute exceptions to this general trend. Retrievals have shown significant discrepancies between particle sizes of mesospheric clouds during the GDS and those during the January 2019 dust event. We have attributed this difference to the larger availability of $H_2O$ and dust in the mesosphere during the GDS than the 2019 event, which results in differing condensation efficiencies.

These elements, together with the observed large vertical and temporal variability of water ice particle size pose questions about the description of water ice nucleation processes into models. In particular, the accuracy of NOMAD retrievals of water ice in the mesosphere up to 100 km constitute a precious source to validate current working hypotheses on the role of both planetary and interplanetary dust as condensation nuclei at various altitudes. Furthermore, the observations we have presented are important to fill in the existing temporal gaps in the literature, and can serve as a database to be assimilated into global circulation models, going beyond the simple elaboration of climatology for these quantities.

**Acknowledgments, Samples, and Data**

ExoMars is a space mission of the European Space Agency (ESA) and Roscosmos. The NOMAD experiment is led by the Royal Belgian Institute for Space Aeronomy (IASB-BIRA), assisted by Co-PI teams from Spain (IAA-CSIC), Italy (INAF-IAPS), and the United Kingdom (Open University). This project acknowledges funding by the Belgian Science Policy Office (BELSPO), with the financial and contractual coordination by the ESA Prodex Office (PEA 4000103401, 4000121493), by the Spanish MICINN through its Plan Nacional and by European funds under grants PGC2018-101836-B-I00 and ESP2017-87143-R (MINECO/FEDER), as well as by UK Space Agency through grants ST/R005761/1, ST/P001262/1, ST/R001405/1 and ST/S00145X/1 and Italian Space Agency through grant 2018-2-HH.0. The IAA/CSIC team acknowledges financial support from the State Agency for Research of the Spanish MCIU through the "Center of Excellence Severo Ochoa" award for the Instituto de Astrofísica de Andalucía (SEV-2017-0709). This work was supported by NASA's Mars Program Office under WBS 604796, "Participation in the TGO/NOMAD Investigation of Trace Gases on Mars" and by NASA's SEEC initiative under Grant Number NNX17AH81A, "Remote sensing of Planetary Atmospheres in the Solar System and Beyond". MC is supported by the NASA Postdoctoral Program at the NASA Goddard Space Flight Center, administered by Universities Space Research Association (USRA) under contract with NASA.

The retrieval package used in this study is the Planetary Spectrum Generator, free and available online at https://psg.gsfc.nasa.gov/helpatm.php#retrieval. The database with the retrieved values are available of the PSG Exomars server at https://psg.gsfc.nasa.gov/apps/exomars.php.